\documentclass[doublecol]{epl2}

\usepackage{bm,graphicx,graphics,amsmath,amssymb,bm,color,pifont}
\usepackage{euscript,tabularx,textcomp}
\usepackage[english]{babel}

\usepackage{psfig}

\begin{document}

\title{Are superparamagnetic spins classical?}
\shorttitle{Are superparamagnetic spins classical?}
\author{D. A. Garanin}
\shortauthor{D. A. Garanin}

\institute{
  Physics Department, Lehman College, City University of New York, 250 Bedford Park Boulevard West, Bronx, New York 10468-1589, U.S.A.\\
}
\pacs{75.50.Tt}{Fine particle systems}
\pacs{75.10.Hk}{Classical spin models}
\pacs{75.50.Xx}{Molecular magnets}

\date{\today}

\abstract{
Effective giant spins of magnetic nanoparticles are considered classically
in the conventional theory of superparamagnetism based on the
Landau-Lifshitz-Langevin equation. However, microscopic calculations for a
large spin with uniaxial anisotropy, coupled to the lattice via the simplest
generic mechanism, show that the results of the conventional theory are not reproduced in the
limit $S\rightarrow \infty .$ In particular, the prefactor $\Gamma _{0}$ in
the Arrhenius escape rate over the barrier $\Gamma =\Gamma _{0}\exp \left[
-\Delta U/\left( k_{\mathrm{B}}T\right) \right] $ has an anomalously large
sensitivity to symmetry-breaking interactions such as transverse field.
}

\maketitle

Ferromagnetic particles of a sufficiently small size (e.g., magnetic
nanoparticles) are in a single-domain magnetic state, atomic spins being
kept collinear by a strong exchange interaction. The resulting giant spin of
a magnetic particle shows a bistability in the case of uniaxial anisotropy
that creates two energy minima and a barrier between them \cite
{stowoh48,stowoh91}. At thermal equilibrium, there is a distribution over
directions of particles' spins similar to that of paramagnets. Since total
spins of magnetic particles are very large, this kind of paramagnetism is
called ``superparamagnetism''.

N\'{e}el suggested a model of relaxation of ensembles of magnetic particles
in which spins are hopping between the two energy minima \cite{nee49}.
Modern approach to superparamagnetic dynamics is based on the
Landau-Lifshitz equation \cite{lanlif35} for classical spins of fixed length
augmented by the stochastic Langevin field simulating the environment \cite
{bro63pr} that is equivalent to the Fokker-Planck equation (FPE) for
classical spins. Solution of the FPE yields the Arrhenius thermal activation
rate $\Gamma =\Gamma _{0}\exp \left[ -\Delta U/\left( k_{\mathrm{B}}T\right) %
\right] $ for $T\ll \Delta U/k_{\mathrm{B}},$ $\Delta U$ being the energy
barrier \cite{bro63pr,aha64pr}.

The amount of theoretical papers published on the subject up to now is
innumerable. The reader can refer to the book \cite{cofcalwal96book} on the
Langevin approach to magnetic and dipolar systems and to Ref. \cite
{cofgarcar01acp} for a review of spin thermal activation problems.
Numerically one can solve the FPE using matrix continued fractions \cite
{cofcrokalwal95prb} or other methods. Alternatively, one can start with the
underlying stochastic model and solve it with matrix continued fractions
\cite{titkackalcof05prb} or directly as a stochastic differential equation
\cite{libcha93jap,pallaz98prb}, also for the model with a variable spin
length near the Curie temperature \cite{garfes04prb}. For a model of many
classical atomic spins forming a nanoparticle, direct solution of a system
of Landau-Lifshitz-Langevin (LLL) equations is the only working numerical
method \cite{chunowchagar06prb}.

With increasing the size of a magnetic particle, noncollinearities due to
the surface anisotropy proliferate and the single-domain state is gradually
destroyed. Small noncollinearities result in an additional cubic anisotropy for
particle's global magnetization \cite{garkac03prl,kacbon06prb,yanetal07prb}
that is second-order in the surface-anisotropy constant and scales as the
particle's volume. Another interesting effect is relaxation via internal spin waves \cite
{safber01prb,garkacrey08epl} that should add up with the relaxation due to the
environment.

Experimentally, only recent successes in fabrication of nanoparticles with
well-controlled parameters allowed to obtain the famous Stoner-Wohlfarth
astroid and check the N\'{e}el-Brown theory of thermal activation \cite
{weretal97prl,cofetal98prl,bonetal99prl}. On smaller nanoparticles,
indications of spin tunneling \cite
{chu79jetp,enzsch86,chugun88prl,chutej98book} have been seen \cite
{weretal97prl-q}. Later, however, the interest in spin tunneling has shifted
to molecular magnets such as Mn$_{12}$ and Fe$_{8}$, where the molecular
spin is only $S=10$ and the phenomenon could be observed with a much greater
certainty and resolution \cite
{frisartejzio96prl,heretal96epl,thoetal96nat,werses99science,weretal00epl}.

The stochastic model of magnetic particles using the Landau-Lifshitz
equation for a large spin with the formal Langevin magnetic field has been
perpetuated in the literature because of its simplicity. However, this model
contradicts the time-reversal symmetry. Deformations of the lattice due to
thermal fluctuations cannot produce any fluctuating effective magnetic field
(i.e., terms in the Hamiltonian \emph{linear} in spin components). \ It
rather produces a fluctuating anisotropy, i.e., stochastic terms \emph{even}
in components of the spin $S.$ The corresponding analysis has been done in
Ref. \cite{garishpan90tmf} where it was shown how the symmetry and strength
of the relaxation term in the Landau-Lifshitz equation follows from those of
the stochastic terms. However, this model for classical spins was never used
because it can include too many difficult-to-define coupling and damping constants.

On the other hand, all microscopic quantum-mechanical models of spin-lattice
relaxation employ spin-lattice couplings that do not violate basic
symmetries. Until recently, however, these calculations suffered from too
many unknown coupling constants that allowed only order-of-magnitude
estimations. Discovery of the universal mechanism of spin relaxation via
distortionless rotation of the crystal field by transverse phonons \cite
{chu04prl,chugarsch05prb} changed the situation. Within this mechanism,
spin-lattice coupling can be expressed through the parameters of the crystal
field that can be easily measured. Implementing this mechanism in the
stochastic formalism for classical spins of Ref. \cite{garishpan90tmf} would
allow to rewrite the theory of superparamagnetism in a more satisfactory
form.

The less ambitious aim of this Letter is, however, just to demonstrate that relaxation of large spins of
magnetic particles cannot be described by the conventional classical approach. The point
is that the most important quantum-mechanical relaxation processes such as
emission/absorption of phonons are sensitive to the energy levels of the
spin. Since in the existing LLL formalism the information of the
energy levels is lost, there is no connection to the underlying
quantum mechanics and the ensuing results are questionable.

One can argue that giant spins of magnetic particles, $S\ggg 1,$ are
classical to a high degree of precision. This is not true, however, since
even the relaxation in the bulk is governed by quantum mechanics. Of course,
equilibrium properties of superparamagnets are classical since one has the
Langevin function instead of the Brillouin function for the field-dependent
magnetization. The relaxation remains non-classical, however, whatever large
is the particle. Indeed, quantum effects in magnetic particles have been
recently observed and discussed in Ref. \cite{nogetal08prb}.

To understand the importance of quantum effects in magnetic particles, one
has to realize the difference between the \emph{classical-spin limit} and
the \emph{large-spin limit}. The classical-spin limit is a theoretical trick
to simplify calculations by eliminating quantum effects. The large-spin
limit, to the contrary, is the real situation.

For instance, for a system of $N$ atomic spins $s$ with the easy-axis
Hamiltonian $\hat{H}=-ds_{z}^{2},$ held together by a strong exchange, the
total spin is $S=Ns\ggg 1.$ Within the classical-spin limit, the effective
classical Hamiltonian of the system would be $\mathcal{H}=-DS_{z}^{2}$ with $%
D=d/N^{2}$ to preserve the energy barrier $\Delta U=ds^{2}=DS^{2}.$ With the
energy levels of the spin $S$ given by $\varepsilon _{m}=-Dm^{2},$ the
transition frequency $\omega _{S-1,S}=\varepsilon _{S-1}-\varepsilon _{S}$
between the ground and first excited states of the effective spin becomes $%
\hbar \omega _{S,S-1}=\left( 2S-1\right) D\cong 2sd/N$, disappearing in the
limit $N\rightarrow \infty .$ Accordingly, the direct phonon processes die
out for large $N,$ so that the only relaxation processes due to phonons
become the two-phonon Raman processes that become insensitive to the energy
levels for small transition frequencies.

In the realistic large-spin limit, the transition frequency $\omega _{S,S-1}$
is preserved, because this is the frequency of the small-amplitude spin
precession in the anisotropy field. Thus the effective anisotropy constant $%
D $ scales according to $\hbar \omega _{S,S-1}=\left( 2S-1\right) D\cong
2SD=2NsD=2sd=\mathrm{const},$ hence $D=d/N.$ In this case the energy barrier
is $\Delta U=DS^{2}=Nds^{2}\varpropto N,$ the size of the particle, as it
should be. One can see that direct spin-phonon processes survive in the
large-spin limit. This makes the situation completely different from the
classical-spin limit, regarding the relaxation.

On the other hand, the transition frequencies between the levels near the
top of the barrier, $m\sim 1,$ are of order $\hbar \omega _{m,m-1}=\left(
2m-1\right) D\sim D\varpropto 1/N$ and they vanish in the large-spin limit.
This means that direct phonon processes between the adjacent evergy levels,
having the rate $\Gamma _{m.m-1}^{(1)}\varpropto \omega _{m,m-1}^{2}$ for $%
\hbar \omega _{m,m-1}\ll k_{\mathrm{B}}T,$ die out near the top of the
barrier that becomes a bottleneck for the thermal activation process. In
this region, diffusion of spin populations over the stairway of adjacent
levels is effectuated by much weaker Raman processes that leads to small
escape prefactors $\Gamma _{0}$ with essential temperature dependence \cite
{gar97pre}.

Transverse magnetic field $H_{\perp }$ or transverse anisotropy create
saddles in the potential landscape of the effective spin that strongly
change dynamics of thermal activation. A ``phase diagram'' of different
regimes, such as uniaxial, high-, intermediate-, and low-damping regimes,
created by the transverse field, has been obtained in Ref. \cite
{garkencrocof99pre}. Especially in the low-damping (LD) case, transverse
field results in a strong increase of the escape rate $\Gamma .$ As can be
seen from the comparison of the LD and HD cases in Fig.\ 3 of Ref. \cite
{garkencrocof99pre}, the main effect is the increase of the prefactor $%
\Gamma _{0},$ while lowering the barrier $\Delta U$ (equal in the LD and HD
cases) plays a secondary role.

For a quantum large spin, the effect of transverse field $H_{\perp }$ should
be even greater, since for $H_{\perp }=0$ the prefactor $\Gamma _{0}$ is
anomalously small. For $H_{\perp }\neq 0,$ the states $\left| m\right\rangle
$ are no longer eigenvalues of the spin Hamiltonian $\mathcal{H,}$ and spin
hopping is no longer restricted to ajacent levels. Thus transverse field
should resolve the bottleneck near the top of the barrier, leading to a huge
increase of the escape prefactor $\Gamma _{0}.$ The aim of the present work
is to describe this effect by solving the density matrix equation (DME) that
is a quantum counterpart of the FPE. The universal mechanism of spin-lattice
relaxation \cite{chu04prl,chugarsch05prb} has been recently incorporated
into the DME \cite{gar08-DME}. Here it will be used to obtain the results
with only one parameter describing the spin-phonon interaction, the
characteristic energy $E_{t}\equiv \left( \rho v_{t}^{5}\hbar ^{3}\right)
^{1/4},$ where $\rho $ is the mass density of the lattice and $v_{t}$ is the
speed of transverse sound. The model used here is on the same level of
simplicity as the standard Landau-Lifshitz-Langevin equation or the FPE but
it is much better justified. We will see that quantum effects on the thermal
activation rate $\Gamma $ do not vanish and become even stronger in the
large-spin limit $S\ggg 1$ for nearly uniaxial magnetic particles.

The effective-spin Hamiltonian has the form
\begin{equation}
\hat{H}_{S}=\hat{H}_{A}+\hat{H}_{\mathrm{Z}},  \label{DME-MMHam}
\end{equation}
where $\hat{H}_{A}$ is the crystal-field (anisotropy) Hamitonian and $\hat{H}%
_{\mathrm{Z}}$ is the Zeeman Hamiltonian,
\begin{equation}
\hat{H}_{A}=-DS_{z}^{2},\qquad \hat{H}_{\mathrm{Z}}=gm_{\mathrm{B}}\left(
H_{z}S_{z}+H_{x}S_{x}\right) .  \label{HamADef}
\end{equation}
The classical energy barrier $\Delta U$ has particular forms
\begin{equation}
\Delta U=DS^{2}\times \left\{
\begin{array}{cc}
(1-h_{x})^{2}, & h_{z}=0 \\
(1-h_{z})^{2}, & h_{x}=0,
\end{array}
\right.  \label{DeltaUparticular}
\end{equation}
where $h_{x,z}\equiv gm_{\mathrm{B}}H_{x,z}/(2SD).$ In general $\Delta
U(h_{x},h_{z})$ can be visualized as a Stoner-Wohlfarth astropyramid,
completely symmetric in $h_{x}$ and $h_{z}$ and basing on the astroid $%
h_{x}^{2/3}+h_{z}^{2/3}=1.$

In the absence of the transverse field $H_{x},$ the eigenstates of the spin
are $\left| m\right\rangle ,$ $m=-S,\ldots ,S,$ the energy levels being
\begin{equation}
\varepsilon _{m}=-Dm^{2}-gm_{\mathrm{B}}H_{z}m.  \label{DME-epsilonm}
\end{equation}
Condition $\hbar \omega _{mm^{\prime }}\equiv \varepsilon _{m}-\varepsilon
_{m^{\prime }}=0$ for $m\neq m^{\prime }$  defines the resonance values of
the longitudinal field $H_{z}$:
\begin{equation}
gm_{B}H_{z}=kD,\qquad k=0,\pm 1,\pm 2,\ldots  \label{DME-HzRes}
\end{equation}
For these fields \emph{all} levels in the right well $m^{\prime }=-m-k$ are
at resonance with the corresponding levels in the left well, $m<0.$

The magnetic particle can be considered as embedded in the elastic matrix
described by the harmonic-phonon Hamiltonian $\hat{H}_{\mathrm{ph}}=\sum_{%
\mathbf{k}\lambda }\hbar \omega _{\mathbf{k}\lambda }a_{\mathbf{k}\lambda
}^{\dagger }a_{\mathbf{k}\lambda }.$ Approach developed in Refs. \cite
{chu04prl,chugarsch05prb} allows to avoid using unknown spin-phonon coupling
constants and to greatly simplify the formalism. Considering the lattice
locally rotated by transverse phonons without distortion of its crystal
field, one obtains the spin-phonon interaction
\begin{equation}
\hat{H}_{\mathrm{s-ph}}=\hat{R}\hat{H}_{A}\hat{R}^{-1}-\hat{H}_{A},\qquad
\hat{R}=e^{-i\mathbf{S}\cdot \delta \mathbf{\phi }},  \label{DME-HsphR}
\end{equation}
where $\delta \mathbf{\phi }$ is a small rotation angle given by $\delta
\mathbf{\phi =}(1/2)\nabla \times \mathbf{u}(\mathbf{r}),$ $\mathbf{u}(%
\mathbf{r})$ being the lattice displacement due to phonons. Expanding Eq.\ (%
\ref{DME-HsphR}) up to first order in $\delta \mathbf{\phi }$ yields the
spin-phonon interaction that describes one-phonon processes:
\begin{equation}
\hat{H}_{\mathrm{s-ph}}^{(1)}=i\left[ \hat{H}_{A},\mathbf{S}\right] \cdot
\delta \mathbf{\phi .}  \label{DME-V1}
\end{equation}
It is important that the spin-phonon interaction above does not include any
poorly known spin-lattice coupling coefficients and it is entirely
represented by the crystal field $\hat{H}_{A}.$ To describe the two-phonon
(Raman) processes, one has to expand $\hat{H}_{\mathrm{s-ph}}$ up to the
second order in $\delta \mathbf{\phi }$ \cite{calchugar06prb,gar08-DME}. \
Relaxation rates due to Raman processes are generally much smaller than those
due to the direct processes since they are the next order in the spin-phonon
interaction. However, the rates of direct processes can be small for special
reasons, then Raman processes become important. Here it happens indeed near
the top of the barrier in zero transverse field, where the transition
frequencies between adjacent levels become small. This situation has been
studied in detail in Ref. \cite{gar97pre}, however. So we will neglect Raman
proceses here and concentrate on the effect of the transverse field that
change transition frequencies and drastically increase the escape rate.

We use the canonical quantization of the lattice displacement $\mathbf{u}$
that yields
\begin{equation}
\delta \mathbf{\phi }=\frac{1}{2}\sqrt{\frac{\hbar }{2MN}}\sum_{\mathbf{k}%
\lambda }\frac{\left[ i\mathbf{k}\times \mathbf{e}_{\mathbf{k}\lambda }%
\right] e^{i\mathbf{k\cdot r}}}{\sqrt{\omega _{\mathbf{k}\lambda }}}\left(
a_{\mathbf{k}\lambda }+a_{-\mathbf{k}\lambda }^{\dagger }\right) .
\label{DME-deltaphiPhiQuantized}
\end{equation}
Here $M$ is the mass of the unit cell, $N$ is the number of cells in the
crystal, $\mathbf{e}_{\mathbf{k}\lambda }$ are unit polarization vectors, $%
\lambda =t_{1},t_{2},l$ denotes polarization, and $\omega _{k\lambda
}=v_{\lambda }k$ is the phonon frequency. Only transverse phonons, $\mathbf{e%
}_{\mathbf{k}\lambda }\bot \mathbf{k},$ survive in this formula.

Spin-lattice relaxation including thermal activation can be described by the
density-matrix equation (DME) \cite{blu81,gar08-DME}. Early application of
the DME to the present model in Ref. \cite{garchu97prb} used the natural basis of states $\left|
m\right\rangle .$ This provided an overall satisfactory description of the
thermal activation rate, including its strong increase at resonance values
of $H_{z}$ given by Eq.\ (\ref{DME-HzRes}). On the other hand, exact energy
levels $\left| \alpha \right\rangle $ of the spin strongly differ from $%
\left| m\right\rangle $ near the top of the barrier even for a small $H_{x}.$
For this reason, the DME below will be written with respect to the energy
basis $\left| \alpha \right\rangle $ obtained by numerical diagonalization
of $\hat{H}_{S}$ \cite{gar08-DME}.

The relaxation terms in the DME can be represented in the form that does not
explicitly contain $\hat{H}_{A},$ the information about it being absorbed in
the spin eigenstates $\left| \alpha \right\rangle $ and transition
frequencies $\omega _{\alpha \beta }.$ This can be achieved either by
changing from the laboratory frame to the local lattice frame in which $\hat{%
H}_{A}$ remains constant but an effective rotation-generated magnetic field
arises \cite{chutej98book,chu04prl,chugarsch05prb}, or by manipulating
matrix elements of the spin-phonon interaction with respect to exact spin
states, $\left\langle \alpha \left| \hat{H}_{\mathrm{s-ph}}^{(1)}\right|
\beta \right\rangle $ \cite{chugarsch05prb}. Both methods are mathematically
equivalent \cite{chugarsch05prb}. As a result, the spin part of spin-phonon
matrix elements is given by the universal expression
\begin{eqnarray}
\mathbf{\Xi }_{\alpha \beta }^{(1)} &\equiv &i\left\langle \alpha \left| %
\left[ \hat{H}_{A},\mathbf{S}\right] \right| \beta \right\rangle =i\hbar
\omega _{\alpha \beta }\left\langle \alpha \left| \mathbf{S}\right| \beta
\right\rangle  \notag \\
&&\qquad -\left\langle \alpha \left| \mathbf{S}\right| \beta \right\rangle
\mathbf{\times }g\mu _{\mathrm{B}}\mathbf{H}.  \label{DME-XiDef}
\end{eqnarray}

\begin{figure}[t]
\centerline{\includegraphics[angle=-90,width=8cm]{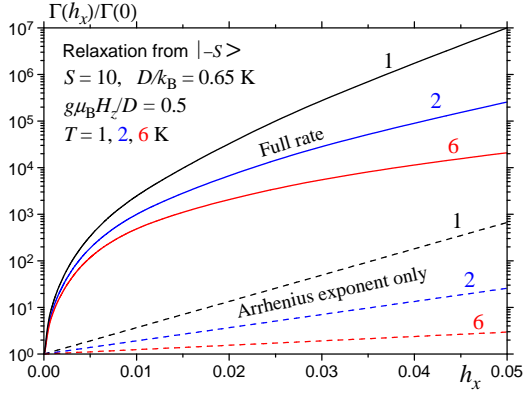}}
\caption{Reduced thermal activation rate vs transverse field for different
temperatures and $S=10$, $D/k_{\mathrm{B}}=0.65$ K. Dramatic increase of $\Gamma $ at small $h_{x}$ is due
to that of the prefactor $\Gamma _{0},$ whereas the activation exponent
(straight dashed lines) gives a comparatively moderate growth. }
\label{Fig-Gamma_S=10}
\end{figure}

At tunneling resonances, Eq.\ (\ref{DME-HzRes}), one has to use the full
non-secular form of the DME that couples diagonal elements of the density
matrix, $\rho _{\alpha \alpha }=n_{\alpha },$ to nondiagonal elements \cite
{gar08-DME}. In the sequel, tunneling resonances will be avoided by choosing
the bias field $H_{z}$ in the middle between the resonances, to make a
better connection with classical models. In this case, one can use the
system of rate equations for the level populations
\begin{equation}
\frac{d}{dt}n_{\alpha }=\sum_{\alpha ^{\prime }=1}^{2S+1}\left( \Gamma
_{\alpha \alpha ^{\prime }}n_{\alpha ^{\prime }}-\Gamma _{\alpha ^{\prime
}\alpha }n_{\alpha }\right) ,  \label{DME-rateEqs}
\end{equation}
where relaxation rates are given by
\begin{eqnarray}
\Gamma _{\alpha \alpha ^{\prime }} &=&2\left( \left| \mathbf{\Xi }_{\alpha
\alpha ^{\prime }}^{(1)}\right| /D\right) ^{2}\left[ \Gamma ^{(1)}\left(
\omega _{\alpha ^{\prime }\alpha }\right) \left( n_{\omega _{\alpha ^{\prime
}\alpha }}+1\right) \right.  \notag \\
&&\qquad +\left. \Gamma ^{(1)}\left( \omega _{\alpha \alpha ^{\prime
}}\right) n_{\omega _{\alpha \alpha ^{\prime }}}\right] .
\label{DME-GammaaaprviaXi2}
\end{eqnarray}
Here $n_{\omega }\equiv \left( e^{\hbar \omega /(k_{\mathrm{B}}T)}-1\right)
^{-1}$ and
\begin{equation}
\Gamma ^{(1)}(\omega )\equiv \frac{\left| \omega \right| ^{3}D^{2}}{24\pi
\hbar ^{2}\Omega _{t}^{4}}\theta (\omega ),  \label{Gamma1}
\end{equation}
$\theta (\omega )$ being a Heavyside function and $\Omega _{t}\equiv \left(
\rho v_{t}^{5}/\hbar \right) ^{1/4}$ being a characteristic frequency. \ In
Eq.\ (\ref{DME-rateEqs}) transitions occur between all the exact spin levels $%
\alpha ,$ although $\Gamma _{\alpha \alpha ^{\prime }}$ corresponding to
pairs of adjacent levels are still dominating. On
the other hand, small transition rates $\Gamma _{\alpha \alpha ^{\prime }}$
near the top of the barrier are strongly modified even for $h_{x}\ll 1.$ The
coupling of the spin to the environment is gauged by a single parameter, $%
\Omega _{t}$ in Eq.\ (\ref{Gamma1}), similarly to the parametrization by the
dimensionless damping constant $\alpha $ in the classical LLL equation.
However, in the present quantum model the rate $\Gamma ^{(1)}(\omega )$ is
frequency dependent through the distances between energy levels that has no
analog in the classical scheme.

Numerical solution of Eq.\ (\ref{DME-rateEqs}) for the parameters of the
molecular magnet Mn$_{12}$ ($S=10$, $D/k_{\mathrm{B}}=0.65$ K) shifted away
from the zero-field resonance, $gm_{B}H_{z}=0.5D,$ shows a huge dependence
on the transverse field $h_{x},$ mainly due to the increase of the prefactor
$\Gamma _{0}$ (see Fig.\ \ref{Fig-Gamma_S=10}). The contribution of the
Arrhenius exponent $\exp \left[ -\Delta U/\left( k_{\mathrm{B}}T\right) %
\right] $ to the growth of $\Gamma (h_{x})$, shown by straight lines $\exp %
\left[ 2h_{x}DS^{2}/\left( k_{\mathrm{B}}T\right) \right] $ following from
Eq.\ (\ref{DeltaUparticular}), becomes important only on the right side of
the plot where the growth of $\Gamma _{0}(h_{x})$ saturates. The effect of
the transverse field here is much greater than in the classical model, the
LD curve in Fig.\ 3 of Ref. \cite{garkencrocof99pre}. Note that in the present model we are
in the uniaxial -- low damping limit since the damping calculated here from
the first principles for realistic $\Omega _{t}$
is much smaller than all other frequency scales.

\begin{figure}
\centerline{\includegraphics[angle=-90,width=8cm]{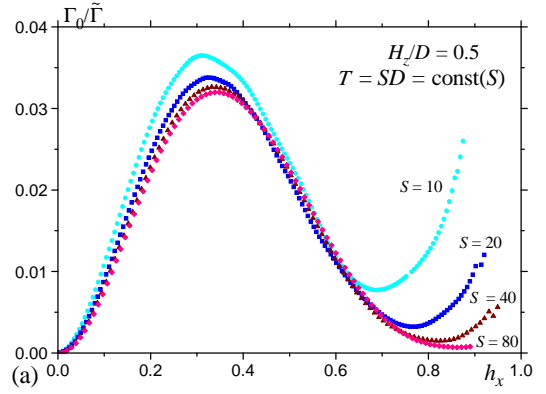}}
\centerline{\includegraphics[angle=-90,width=8cm]{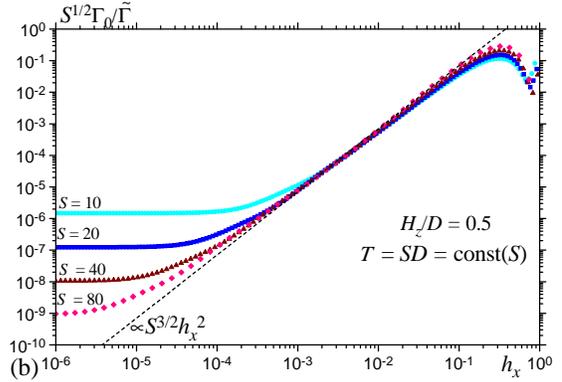}}
\caption{Escape-rate prefactor $\Gamma _{0}$ vs transverse field for
different values of particle's spin $S$ at fixed temperature. }
\label{Fig-Gamma0_nobias}
\end{figure}

For effective spins of magnetic particles that are much greater than $S=10,$
the effect of the transverse field is huge. Since $\Gamma $ in zero
transverse field becomes anomalously small for large spins, one cannot
normalize the results by it. It is better to plot the prefactor $\Gamma _{0}$
alone defined as $\Gamma _{0}=\Gamma \exp \left[ \Delta U/\left( k_{\mathrm{B%
}}T\right) \right] ,$ where $\Gamma $ follows from the solution of Eq.\ (\ref
{DME-rateEqs}) and $\Delta U$ is found numerically for the classical model.
The characteristic rate
\begin{equation}
\tilde{\Gamma}\equiv S\Gamma _{S,S-1}=\frac{S^{2}\omega _{S,S-1}^{5}}{12\pi
\Omega _{t}^{4}}  \label{GammaTil}
\end{equation}
can be used to normalize the results for $\Gamma _{0}$ in a wide range of $%
h_{x}.$ Here $\Gamma _{S,S-1}$ and $\omega _{S,S-1}$ are zero-temperature
relaxation rate and transition frequency for the lowest-lying pair of levels
in the well, defined above. $\tilde{\Gamma}=S\Gamma _{S,S-1}$ is an overall
measure of the relaxation rate inside a well. Indeed, in the natural basis
the spin-phonon transition rate between two adacent levels is proportional
to $\bar{l}_{m,m\pm 1}^{2}$ \cite{garchu97prb}, where $\bar{l}_{m,m\pm
1}\equiv l_{m,m\pm 1}(2m\pm 1)$ and $l_{m,m\pm 1}=\sqrt{S(S+1)-m(m\pm 1)}.$
In $\Gamma _{S,S-1}$ the factor $\bar{l}_{m,m\pm 1}^{2}$ yields $S$ while
for a typical value of $m$ in the interval $-S\leq m\leq S$ it yields $%
S^{2}. $ This is the origin of an additional $S$ in Eq.\ (\ref{GammaTil}).

In the comparizon between different values of $S$ shown in Fig.\ \ref
{Fig-Gamma0_nobias} the product $SD$ is kept constant, as it should be for
the effective anisotropy of magnetic particles. In numerical calculations $%
D/k_{\mathrm{B}}=6.5/S$ K is used and the temperature $k_{\mathrm{B}}T=SD=%
\mathrm{const}.$ Fig.\ \ref{Fig-Gamma0_nobias} (a) shows that curves $\Gamma
_{0}/\tilde{\Gamma}$ scale for large $S$ in a wide range of $h_{x},$ that
means  $\Gamma _{0}\varpropto S^{2}.$ Calculations use custom-precision
matrix algebra within Wolfram Mathematica and become slow for spins as large
as $S=80.$  One can see that in the
large-spin limit $\Gamma _{0}$ becomes small if $h_{x}\rightarrow 0$ and $%
h_{x}\rightarrow 1.$ In particular, for $h_{x}\rightarrow 0$ the apparent
behavior is $\Gamma _{0}\varpropto h_{x}^{2}.$

The behavior of $\Gamma _{0}$ at small transverse fields is elucidated in
Fig.\ \ref{Fig-Gamma0_nobias} (b). Here one has to use a slightly
different normalization of $\Gamma _{0}$ to make curves collapse in a wide
 range of $h_{x}$, yielding $\Gamma
_{0}\varpropto S^{3/2}h_{x}^{2}.$ In the uniaxial limit $h_{x}\rightarrow 0$
the curves for different $S$ diverge. Here the escape prefactor is given by
the transition rate between the ajacent levels near the top of the barrier $%
\Gamma _{m,m\pm 1}$ with $m\simeq 1.$ Using Eq.\ (A9) of Ref. \cite
{chugarsch05prb} for $\Gamma _{m,m\pm 1}$ [multiplied by $n_{\omega _{m,m\pm
1}}\cong k_{\mathrm{B}}T/(\hbar \omega _{m,m\pm 1})$ to account for a
nonzero temperature], one obtains $\Gamma _{0}\varpropto S^{-2}.$ This is
the top-of-the-barrier bottleneck mentioned in the introduction. In the
representation of Fig.\ \ref{Fig-Gamma0_nobias} (b) one has $\Gamma
_{0}/S^{3/2}\varpropto S^{-7/2}.$ One can see that doubling $S$ results in
the drop by a factor $2^{7/2}\simeq 11$ in the asymptotic $h_{x}\rightarrow 0
$ values in Fig.\ \ref{Fig-Gamma0_nobias} (b).
\begin{figure}[t]
\unitlength1cm
\begin{picture}(13,5.5)
\centerline{\psfig{file=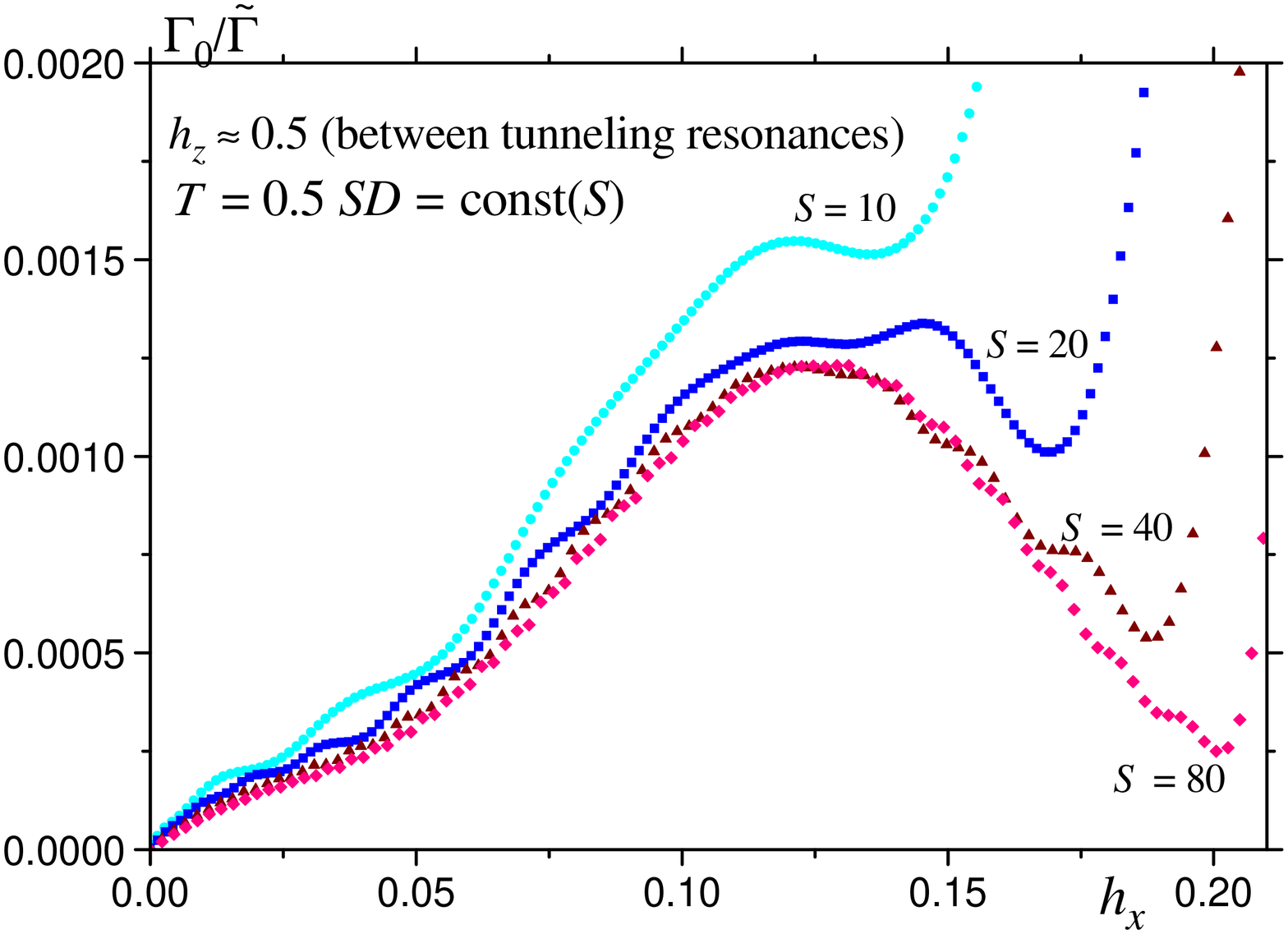,angle=-90,width=8cm}}
\end{picture}
\begin{picture}(13,5.5)
\centerline{\psfig{file=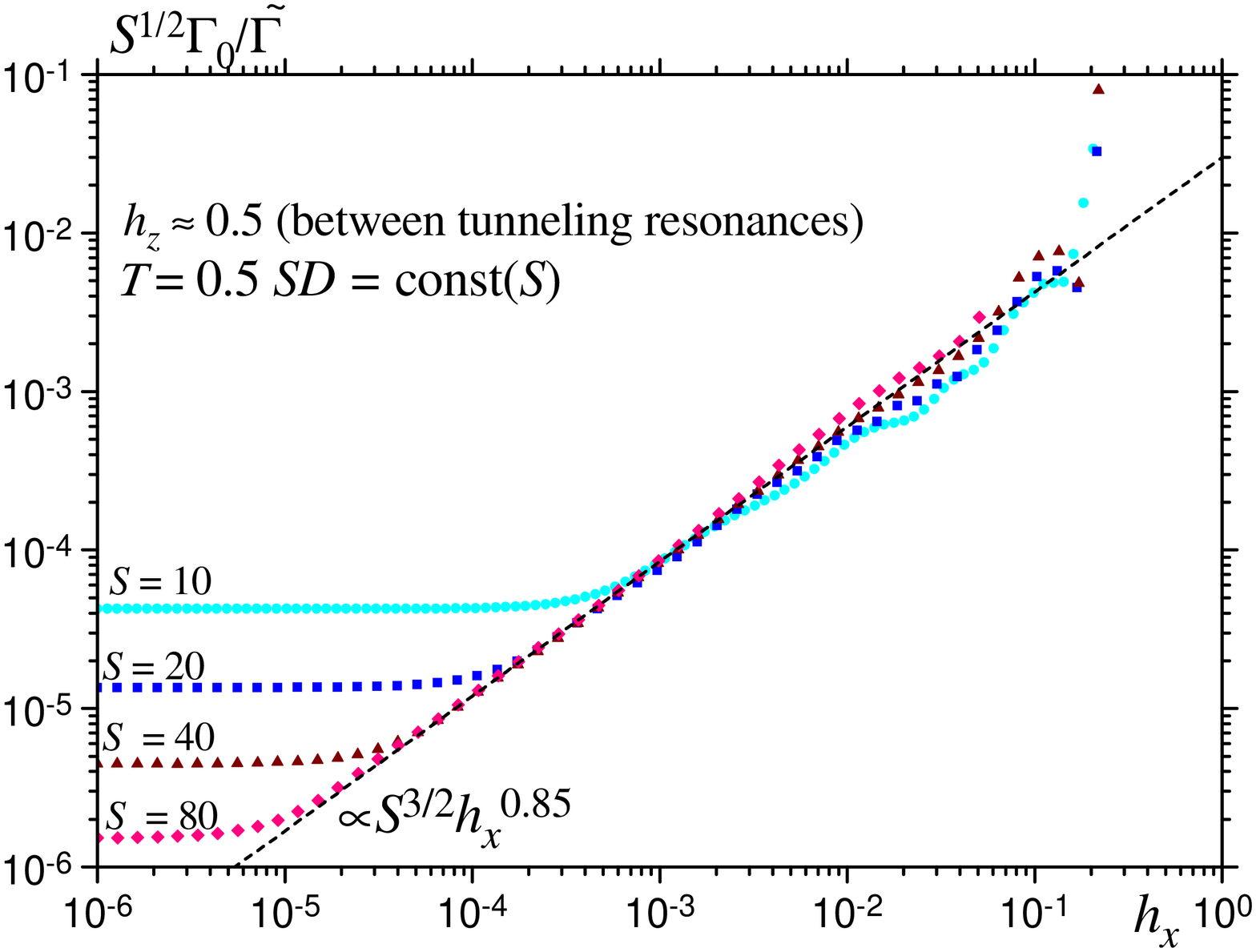,angle=-90,width=8cm}}
\end{picture}
\caption{Escape-rate prefactor $\Gamma _{0}$ vs transverse field for
different values of particle's spin $S$ at fixed temperature at strong bias,
$h_{x}\approx 0.5$ (between tunneling resonances). }
\label{Fig-Gamma0_bias}
\end{figure}

The anomalously small rate in the uniaxial limit above is in part due to the
factor $2m\pm 1$  discussed
below Eq.\ (\ref{GammaTil}). In the zero-bias case the top of the barrier
corresponds to $m\sim 1$ that results in additional smallness. In the case
of a strong enough bias one has  $2m\pm 1\sim S$ near the top of the
barrier, so that the anomalously  small escape rate solely results from
small $\omega _{m,m\pm 1}.$ The results of numerical
calculations for the bias $h_{x}\approx 0.5,$ adjusted to the middle between two tunneling resonances are shown in Fig.\ \ref
{Fig-Gamma0_bias}. The curves for $\Gamma _{0}$ in a
broad range of $h_{x}$ in Fig.\ \ref{Fig-Gamma0_bias} (a) look
complicated for moderate $S$ but still collapse for large $S$. The decrease of $\Gamma _{0}$ at $%
h_{x}\rightarrow 0$ is indeed weaker than in the unbiased case above. The
results at small $h_{x}$ in Fig.\ \ref{Fig-Gamma0_bias} (b) show a dependence
$\Gamma _{0}\varpropto S^{3/2}h_{x}^{0.85},$ where the exponent $0.85$ cannot be
easily explained. For $h_{x}=0$ Eq.\ (A9) of Ref. \cite{chugarsch05prb}
yields $\Gamma _{0}\varpropto S^{0}$ in the biased case, also much smaller than $\Gamma _{0}\varpropto S^{2}$ for $h_{x}\sim 1$.

It should be noted that the secular approximation leading to Eq.\ (\ref
{DME-rateEqs}) relies on the smallness of the relaxation terms in the DME in
comparison to the dissipationless terms for the nondiagonal elements of the
density matrix \cite{gar08-DME}. Then slow diagonal elements $\rho _{\alpha
\alpha }=n_{\alpha }$ dynamically decouple from the fast nondiagonal terms $%
\rho _{\alpha \beta }.$ The classical match of the secular approximation is
the low-damping (LD) approximation introduced by Kramers for a particle in a
potential well \cite{kra40}. In the LD limit the energy of the particle or
spin is nearly conserved, so that the fast motion over constant-energy
trajectories  averages out and what is left is the slow energy
diffusion (see, e.g., Eqs. (15) and (16) of Ref. \cite{garkencrocof99pre}).
Similarly, Eq.\ (\ref{DME-rateEqs}) describe a slow hopping over the quantum
energy levels of the spin.

One can ask whether the richness of damping regimes that exist in the
classical-spin model \cite{garkencrocof99pre} can be realized for a
realistic large quantum spin of a magnetic particle. For instance, the
intermediate-to-large damping (IHD) case requires that the gyroscopic and
relaxation terms in the FPE be comparable. This means that the
dissipationless and relaxation terms in the DME be comparable as well. Of
course, for $S\gg 1$ nondiagonal elements $\rho _{\alpha \beta }$ close to
diagonal become slow as $\omega _{\alpha \beta }.$ However, the relaxation
rate $\Gamma _{\alpha \beta }$ between the states $\alpha $ and $\beta $
scales as $\Gamma _{\alpha \beta }\varpropto \omega _{\alpha \beta }^{2}$ for
$\left| \omega _{\alpha \beta }\right| \ll T$ and decreases faster than $%
\omega _{\alpha \beta }$ in the quasiclassical limit $S\gg 1.$

What can
change the situation is Raman processes that become independent of $\omega
_{\alpha \beta }$ for small $\omega _{\alpha \beta }.$
Incorporating Raman processes requires generalization of the results
of Ref. \cite{gar97pre} for a nonzero transverse field that is a nontrivial
task. As Raman processes are much weaker than direct processes, crossover to
a Raman-dominated behavior requires very large $S$. Although a general nonsecular DME can be solved as
described in Ref. \cite{gar08-DME}, calculations are much slower than those
of Eq.\ (\ref{DME-rateEqs}) and become prohibitive for the required very
large $S.$ For this reason, Raman processes cannot be adequately treated
within this Letter and should be considered elsewhere.

In all cases, even with account of Raman processes, there should be a
bottleneck for spin diffusion near the top of the barrier in the case of
nearly uniaxial magnetic particles. Transverse magnetic field gradually resolves the bottleneck and leads to a huge
increase of the escape-rate prefactor $\Gamma _{0}$ that is more important
than the barrier lowering. This is the main finding of this work.
Thermal activation rates of nearly-uniaxial magnetic particles
 are very sensitive to any deviations from
the axial magnetic symmetry, e.g., due to surface anisotropy \cite{garkac03prl,kacbon06prb,yanetal07prb}. Robust results require a strong enough transverse field.

Whether for large spins the classical stochastic approach could be modified
to embrace the spacing between quantum-mechanical levels that has been shown
to be important, remains an open question. \

This work has been supported by the Cottrell College Science Award of the Research Corporation.

\bibliographystyle{eplbib}
\bibliography{gar-own,chu-own,gar-tunneling,gar-relaxation,gar-oldworks,gar-books,gar-nano-super,gar-surface-nano}

\begin{thebibliography}{10}
\expandafter\ifx\csname url\endcsname\relax\def\url#1{\texttt{#1}}\fi

\bibitem{stowoh48}
\Name{Stoner E.~C. \and Wohlfarth E.~P.} \REVIEW{Philos. Trans. R. Soc. London,
  Ser. A }{240}{1948}{599}.

\bibitem{stowoh91}
\Name{Stoner E.~C. \and Wohlfarth E.~P.} \REVIEW{IEEE Trans. Magn.
  {}{MAG-27}}{1991}{3475}.

\bibitem{nee49}
\Name{N\'{e}el L.} \REVIEW{Ann. Geophys. (C.N.R.S.) }{5}{1949}{99}.

\bibitem{lanlif35}
\Name{Landau L.~D. \and Lifshitz E.~M.} \REVIEW{Phys. Z. Sowjetunion
  }{8}{1935}{153}.

\bibitem{bro63pr}
\Name{W.~F.~Brown J.} \REVIEW{Phys. Rev. }{130}{1963}{1677}.

\bibitem{aha64pr}
\Name{Aharoni A.} \REVIEW{Phys. Rev. }{135A}{1964}{447}.

\bibitem{cofcalwal96book}
\Name{Coffey W.~T., Kalmykov Y.~P. \and Waldron J.~T.} \Book{The {L}angevin
  {E}quation} (Word Scientific, Singapore) 1996.

\bibitem{cofgarcar01acp}
\Name{Coffey W.~T., Garanin D.~A. \and McCarthy D.~J.} \Book{Crossover formulas
  in the {K}ramers theory of thermally activated escape rates - {A}pplication
  to spin sytems} in \Book{Advances in Chemical Physics}, edited by
  \Name{Prigogine I. \and Rice S.~A.} Vol. 117 (John Wiley \& Sons, Inc.) 2001.

\bibitem{cofcrokalwal95prb}
\Name{Coffey W.~T., Crothers D. S.~F., Kalmykov Y.~P. \and Waldron J.~T.}
  \REVIEW{Phys. Rev. B }{51}{1995}{15947}.

\bibitem{titkackalcof05prb}
\Name{Titov S.~V., Kachkachi H., Kalmykov Y.~P. \and Coffey W.~T.}
  \REVIEW{Phys. Rev. B }{72}{2005}{134425}.

\bibitem{libcha93jap}
\Name{Lyberatos A. \and Chantrell R.~W.} \REVIEW{J. Appl. Phys.
  }{73}{1993}{6501}.

\bibitem{pallaz98prb}
\Name{Garc\'ia-Palacios J.~L. \and L\'azaro F.~J.} \REVIEW{Phys. Rev. B
  }{58}{1998}{14937}.

\bibitem{garfes04prb}
\Name{Garanin D.~A. \and Chubykalo-Fesenko O.} \REVIEW{Phys. Rev. B
  }{70}{2004}{212409}.

\bibitem{chunowchagar06prb}
\Name{Chubykalo-Fesenko O., Nowak U., Chantrell R.~W. \and Garanin D.}
  \REVIEW{Phys. Rev. B }{74}{2006}{094436}.

\bibitem{garkac03prl}
\Name{Garanin D.~A. \and Kachkachi H.} \REVIEW{Phys. Rev. Lett.
  }{90}{2003}{065504}.

\bibitem{kacbon06prb}
\Name{Kachkachi H. \and Bonet E.} \REVIEW{Phys. Rev. B }{73}{2006}{224402}.

\bibitem{yanetal07prb}
\Name{Yanes R., Chubykalo-Fesenko O., Kachkachi H., Garanin D.~A., Evans R.
  \and Chantrell R.~W.} \REVIEW{Phys. Rev. B }{76}{2007}{064416}.

\bibitem{safber01prb}
\Name{Safonov V.~L. \and Bertram H.~N.} \REVIEW{Phys. Rev. B
  }{63}{2001}{094419}.

\bibitem{garkacrey08epl}
\Name{Garanin D.~A., Kachkachi H. \and Reynaud L.} \REVIEW{Europhys. Lett.
  }{82}{2008}{17007}.

\bibitem{weretal97prl}
\Name{Wernsdorfer W., Orozco E.~B., Hasselbach K., Benoit A., B.~Barbara N.~D.,
  Loiseau A. \and Mailly D.} \REVIEW{Phys. Rev. Lett. }{78}{1997}{1791}.

\bibitem{cofetal98prl}
\Name{Coffey W.~T., Crothers D. S.~F., Dormann J.~L., Kalmykov Y.~P., Kennedy
  E.~C. \and Wernsdorfer W.} \REVIEW{Phys. Rev. Lett. }{80}{1998}{5655}.

\bibitem{bonetal99prl}
\Name{Bonet E., Wernsdorfer W., Barbara B., Beno\^it A., Mailly D. \and
  Thiaville A.} \REVIEW{Phys. Rev. Lett. }{83}{1999}{4188}.

\bibitem{chu79jetp}
\Name{Chudnovsky E.~M.} \REVIEW{JETP }{50}{1979}{1035}.

\bibitem{enzsch86}
\Name{Enz M. \and Schilling R.} \REVIEW{J. Phys. C }{19}{1986}{L711}.

\bibitem{chugun88prl}
\Name{Chudnovsky E.~M. \and Gunther L.} \REVIEW{Phys. Rev. Lett.
  }{60}{1988}{661}.

\bibitem{chutej98book}
\Name{Chudnovsky E.~M. \and Tejada J.} \Book{Macroscopic quantum tunneling of
  the magnetic moment} (Cambridge University Press, Cambridge) 1998.

\bibitem{weretal97prl-q}
\Name{Wernsdorfer W., Bonet~Orozco E., Hasselbach K., Benoit A., Mailly D.,
  Kubo O., Nakano H. \and Barbara B.} \REVIEW{Phys. Rev. Lett.
  }{79}{1997}{4014}.

\bibitem{frisartejzio96prl}
\Name{Friedman J.~R., Sarachik M.~P., Tejada J. \and Ziolo R.} \REVIEW{Phys.
  Rev. Lett. }{76}{1996}{3830}.

\bibitem{heretal96epl}
\Name{Hern\'andez J.~M., Zhang X.~X., Luis F., Bartolom\'e J., Tejada J. \and
  Ziolo R.} \REVIEW{Europhys. Lett. }{35}{1996}{301}.

\bibitem{thoetal96nat}
\Name{Thomas L., Lionti F., Ballou R., Gatteschi D., Sessoli R. \and Barbara
  B.} \REVIEW{Nature }{383}{1996}{145}.

\bibitem{werses99science}
\Name{Wernsdorfer W. \and Sessoli R.} \REVIEW{Science }{284}{1999}{133}.

\bibitem{weretal00epl}
\Name{Wernsdorfer W., Sessoli R., Caneschi A., Gatteschi D. \and Cornia A.}
  \REVIEW{Europhys. Lett. }{50}{2000}{552}.

\bibitem{garishpan90tmf}
\Name{Garanin D.~A., Ishchenko V.~V. \and Panina L.~V.} \REVIEW{Teor. Mat. Fiz.
  }{82}{1990}{242}.

\bibitem{chu04prl}
\Name{Chudnovsky E.~M.} \REVIEW{Phys. Rev. Lett. }{92}{2004}{120405}.

\bibitem{chugarsch05prb}
\Name{Chudnovsky E.~M., Garanin D.~A. \and Schilling R.} \REVIEW{Phys. Rev. B
  }{72}{2005}{94426}.

\bibitem{nogetal08prb}
\Name{Noginova N., Weaver T., Giannelis E.~P., Bourlinos A.~B., Atsarkin V.~A.
  \and Demidov V.~V.} \REVIEW{Phys. Rev. B }{77}{2008}{014403}.

\bibitem{gar97pre}
\Name{Garanin D.~A.} \REVIEW{Phys. Rev. E }{55}{1997}{2569}.

\bibitem{garkencrocof99pre}
\Name{Garanin D.~A., Kennedy E., Crothers D. S.~F. \and Coffey W.~T.}
  \REVIEW{Phys. Rev. E }{60}{1999}{6499}.

\bibitem{gar08-DME}
\Name{Garanin D.~A.} \REVIEW{arXiv:0805.0391 }{}{2008}{}.

\bibitem{calchugar06prb}
\Name{Calero C., Chudnovsky E.~M. \and Garanin D.~A.} \REVIEW{Phys. Rev. B
  }{74}{2006}{094428}.

\bibitem{blu81}
\Name{Blum K.} \Book{Density {M}atrix {T}heory and {A}pplications} (Plenum
  Press, New York, London) 1981.

\bibitem{garchu97prb}
\Name{Garanin D.~A. \and Chudnovsky E.~M.} \REVIEW{Phys. Rev. B
  }{56}{1997}{11102}.

\bibitem{kra40}
\Name{Kramers H.~A.} \REVIEW{Physica (Amsterdam) }{7}{1940}{284}.

\end{thebibliography}

\end{document}